\documentclass[final,3p,times]{elsarticle}
\usepackage[english]{babel}
\usepackage[T1]{fontenc}
\usepackage{graphicx}
\usepackage{subfigure}
\usepackage{amssymb,amsmath,bm}
\usepackage[colorlinks=true]{hyperref}
\usepackage{booktabs}
\usepackage{multirow}
\biboptions{sort&compress}

\newcommand{\goodgap}{%
	\hspace{\subfigtopskip}%
	\hspace{\subfigbottomskip}}

\journal{Nuclear Physics A}

\begin{document}

\begin{frontmatter}

\title{Auxiliary Field Diffusion Monte Carlo study of the Hyperon-Nucleon interaction in $\Lambda$-hypernuclei}

\author[1]{Diego Lonardoni}
\ead{lonardoni@science.unitn.it}
\address[1]{Dipartimento di Fisica, Universit\`a di Trento and INFN, Gruppo Collegato di Trento, via Sommarive 14, I-38123 Trento, Italy}

\author[1]{Francesco Pederiva}
\ead{pederiva@science.unitn.it}

\author[2]{Stefano Gandolfi}
\ead{stefano@lanl.gov}
\address[2]{Theoretical Division, Los Alamos National Laboratory, Los Alamos, 87545 US-NM}

\begin{abstract}
We investigate the role of two- and three-body $\Lambda$-nucleon forces
by computing the ground state of a few $\Lambda$-hypernuclei with the  
Auxiliary Field Diffusion Monte Carlo algorithm.
Calculations have been performed for masses up to $A=41$, 
including some open-shell hypernuclei. 
The results show that the use of a bare hyperon-nucleon force fitted 
on the available scattering data yields a 
consistent overestimate of the $\Lambda$-separation energy $B_\Lambda$.
The inclusion of a hyperon-nucleon-nucleon interaction systematically reduces $B_\Lambda$,
leading to a qualitatively good agreement with experimental data over the range of
masses investigated.
\end{abstract}

\begin{keyword}
$\Lambda N$ interaction \sep $\Lambda$-hypernuclei \sep
$\Lambda$-separation energy \sep Auxiliary Field Diffusion Monte Carlo
(AFDMC)
\end{keyword}

\end{frontmatter}

\section{Introduction}
\label{sec:intro}
The outer core of a neutron star (NS), where matter is
supposed to have a density close to the saturation value, can be
safely modeled as a gas of neutrons with a small fraction of protons
and leptons (see for example Ref.~\cite{Gandolfi:2010}).
At the larger densities reached in the inner core, the composition of
matter is instead rather uncertain. The available observations of 
NS masses of order 2$M_\odot$~\cite{Demorest:2010,Steiner:2012}
has greatly reduced the softening of the
equation of state (EOS) needed to yield a maximum mass in agreement with 
observational data. This fact seems to strongly question one of the potential sources of 
softening, i.e. the onset of heavier hadrons, such as hyperons,
for large enough values of the chemical potential of NS matter.
Most recent calculations on neutron matter with the
inclusion of hyperons give a rather strong softening of the EOS~
\cite{Dapo:2010,Vidana:2011,Schulze:2011,Massot:2012,Miyatsu:2012}.
However, results are strictly model dependent. A number of hyperon-nucleon
interactions have been used (Nijmegen, J\"ulich, $\chi$EFT), with
no overall agreement in the results. Some other approaches exist, 
that instead predict very small changes to the neutron star structure 
once the hyperons are included in the model \cite{Bednarek:2012,Weissenborn:2012}.

The only reliable benchmarks on
hyperon-nucleon model forces can be made on the experiments with hypernuclei.
Most of the limited available data concern single $\Lambda$-hypernuclei.
Only a few $\Lambda$-nucleon scattering data are available. 
Binding energies, excitation energies and hyperon-separation energies for
approximately 40 $\Lambda$-hypernuclei are available (for recent results see e.g.~\cite{Agnello:2012_H6L,Nakamura:2012_He7L,Agnello:2012_He9L}).
The lack of data reflects in the large arbitrariness in the definition
of a hyperon-nucleon potential.

In this paper we focus on a realistic $\Lambda$-nucleon
potential based on the Argonne-like interaction proposed by Bodmer,
Usmani and collaborators (\cite{Usmani:1995,Usmani:2008} and references
therein). The potential is projected in coordinate space 
including both the $\Lambda N$ the and $\Lambda NN$ channel,
and a realistic description of the hard-core repulsion between baryons.
The final target will be refitting the parameters of the hyperon-nucleon-nucleon
force in order to reproduce the experimental
data on the $\Lambda$-separation energy on a set of selected
hypernuclei in a rather wide mass interval. 

The $\Lambda$ separation energy $B_\Lambda$, defined as the difference
between the binding energy of the nucleus $^{A-1}$Z and that of the corresponding hypernucleus
$^{A\,}_{\,\Lambda}$Z, is here computed by means of the Auxiliary Field
Diffusion Monte Carlo (AFDMC) method \cite{Schmidt:1999}.

\section{Hamiltonians and the $\Lambda$-nucleon interaction}
\label{sec:potentials}
We describe nuclei and $\Lambda$-hypernuclei as systems made of non-relativistic 
point-like particles interacting via the following Hamiltonians:
\begin{equation}
H_{nuc}=\sum_{i=1}^{A-1}\frac{p_i^2}{2m_N}+\sum_{i<j}^{A-1}v_{ij} \,,\quad\quad 
H_{hyp}=H_{nuc}+\frac{p_\Lambda^2}{2m_\Lambda}+\sum_{i=1}^{A-1}v_{\Lambda i}+\sum_{i<j}^{A-1}v_{\Lambda ij} \,,
\end{equation}
where $A$ is the total number of baryons, nucleons plus
the $\Lambda$ particle. In the nucleon Hamiltonian we use the Argonne AV4'
and AV6' two body-potentials \cite{Wiringa:2002}, that are simplified versions
of the Argonne AV18 \cite{Wiringa:1995}. As a test of the sensitivity
of $B_\Lambda$ on the choice of the $NN$ interaction, we also performed
calculations with the Minnesota potential \cite{Thompson:1977}.
The $\Lambda$-nucleon interaction is here modeled  by the two- and three-body
potentials developed by Usmani et al. \cite{Usmani:1995,Usmani:2008}. It
is important to note that the cited nuclear potentials do not provide
the same accuracy as AV18 in fitting $NN$ scattering data. In addition,
three-body $NNN$ forces are purposedly disregarded for technical reasons
related to the AFDMC algorithm used.
We rely on the assumption that when taking the difference
between the binding energies of a nucleus and the 
corresponding $\Lambda$-hypernucleus, most of the
nucleon interaction energy cancels. 
We shall see that this assumption is consistent with our results,
thereby confirming that the specific choice of the nucleon Hamiltonian 
should not significantly affect the results on $B_\Lambda$.

\begin{figure}[ht]
	\begin{center}
		\subfigure[\label{fig:LN}]%
		 	{\includegraphics[height=3.2cm]{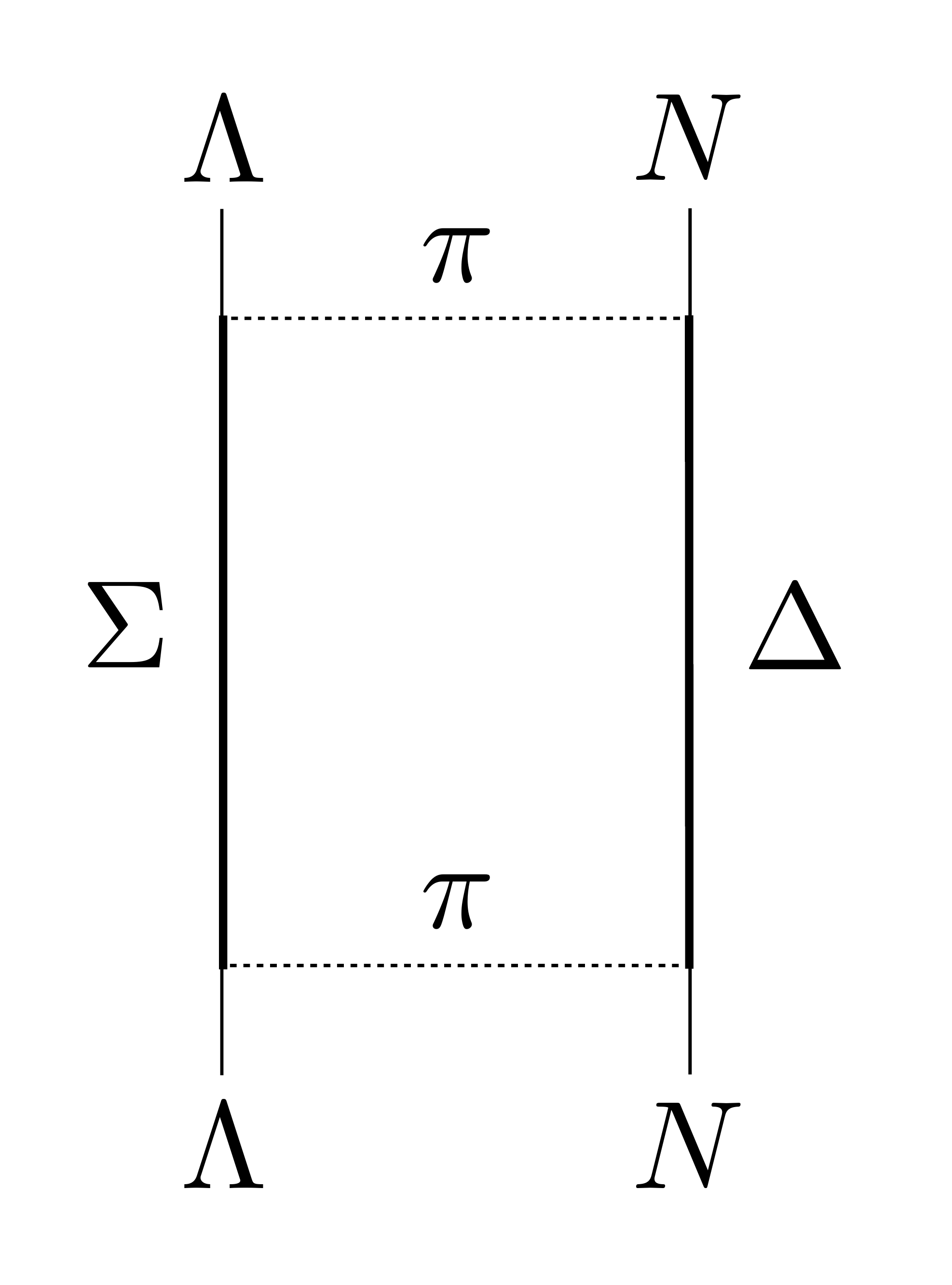}}
		\goodgap\goodgap\goodgap
		\subfigure[\label{fig:LNN_sw}]%
		 	{\includegraphics[height=3.2cm]{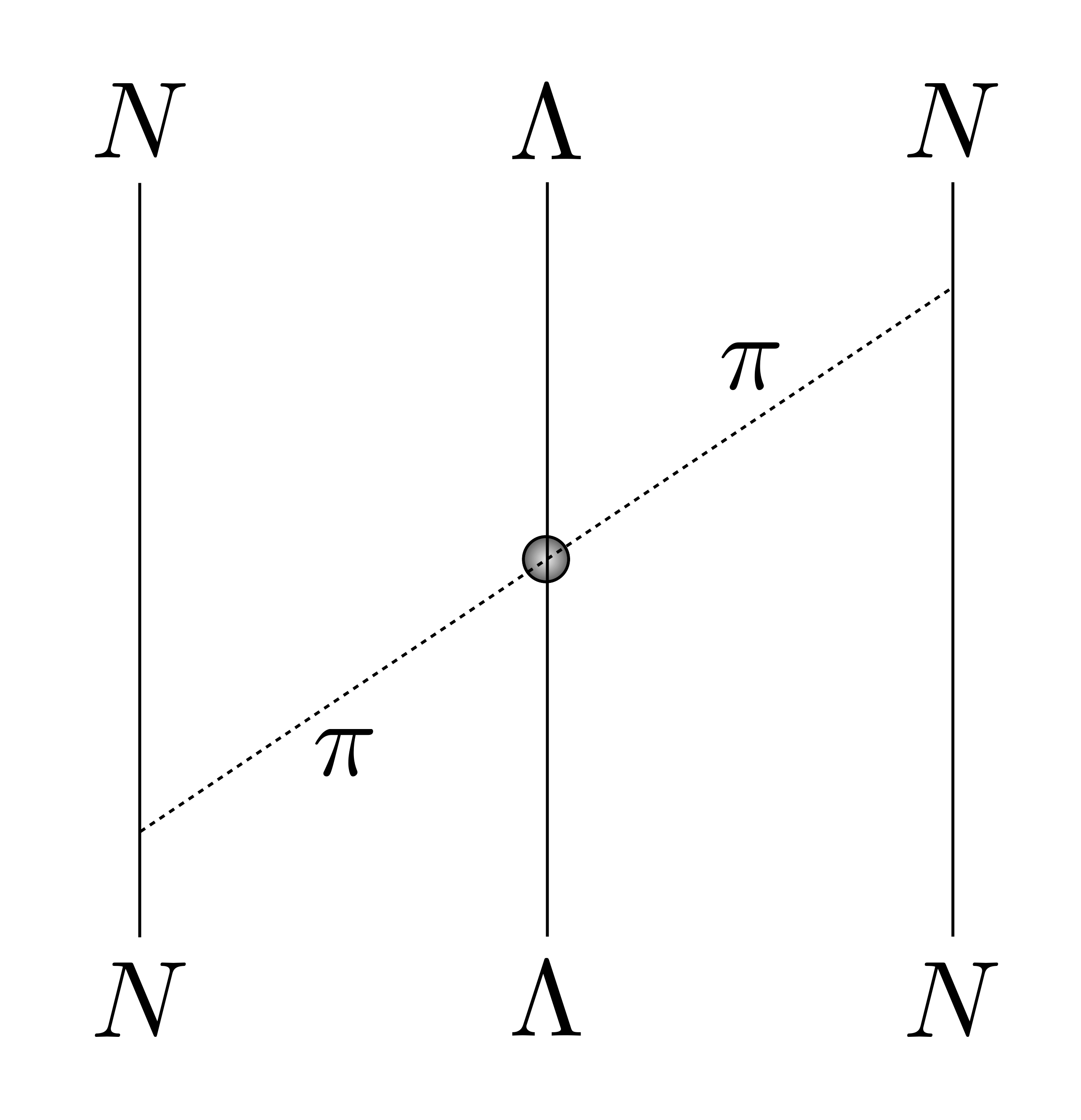}}
		\goodgap\goodgap\goodgap
		\subfigure[\label{fig:LNN_pw}]%
		 	{\includegraphics[height=3.2cm]{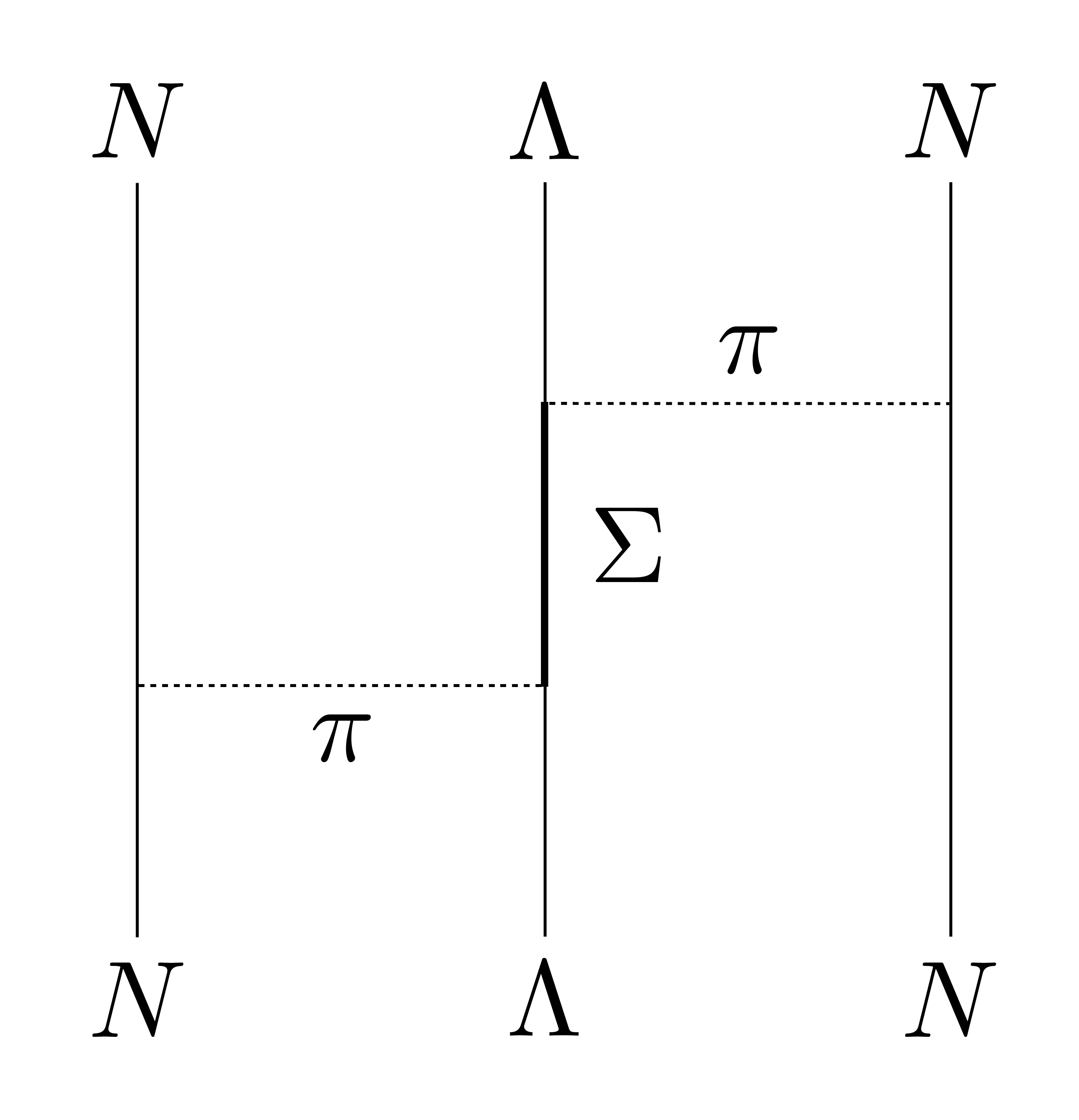}}
		\goodgap\goodgap\goodgap
		\subfigure[\label{fig:LNN_d}]%
		 	{\includegraphics[height=3.2cm]{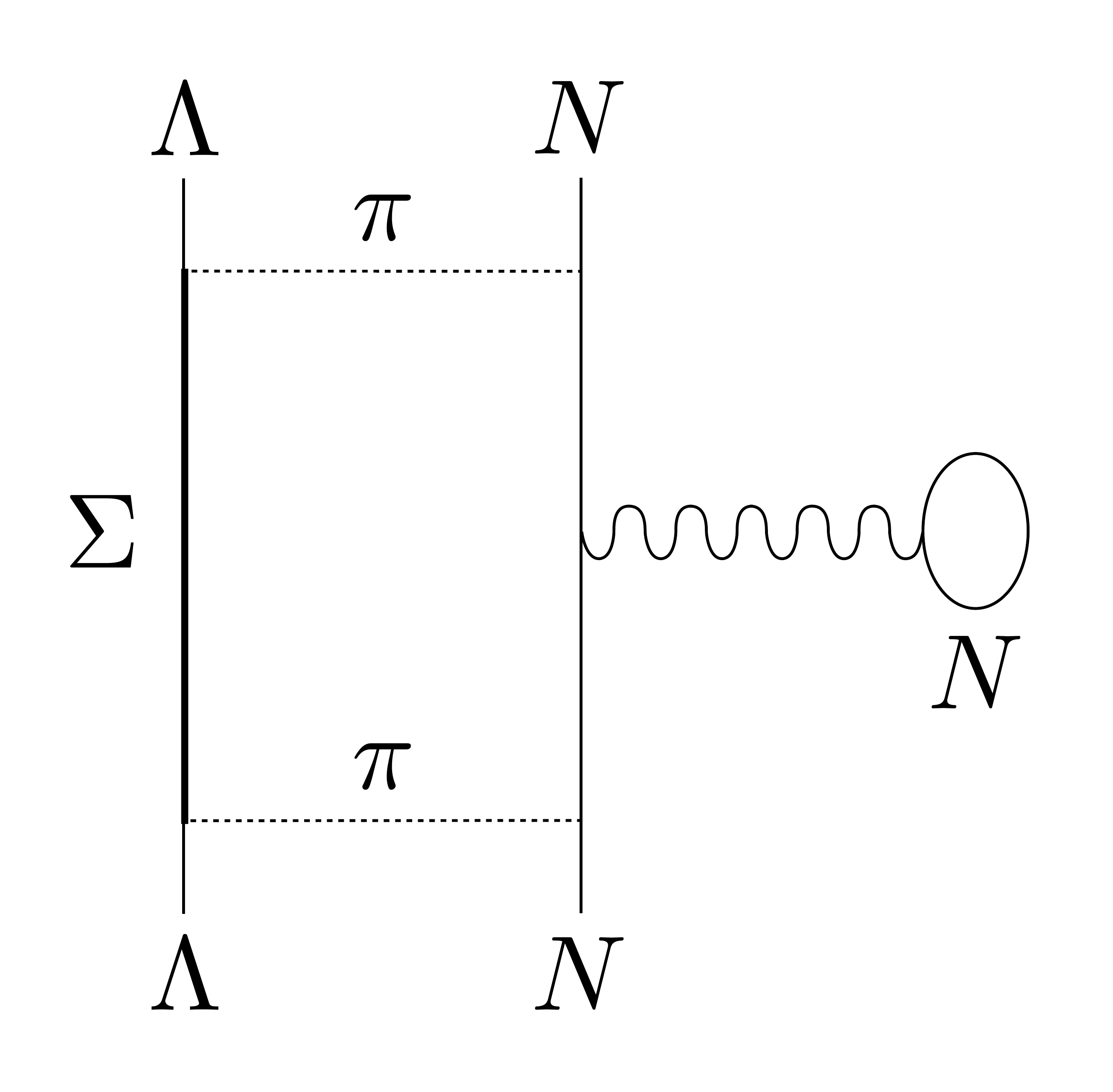}}
	\caption{Schematic construction of the $\Lambda$-nucleon interaction. 
	2$\pi$ exchange processes between nucleons and the
	$\Lambda$ particle. \ref{fig:LN} represents the two-body $\Lambda
	N$ channel. \ref{fig:LNN_sw}, \ref{fig:LNN_pw} and \ref{fig:LNN_d}
	are the three-body $\Lambda NN$ channels. For more details see
	Refs.~\cite{Usmani:1995,Usmani:2008}.}
	\label{fig:LN-LNN} 
	\end{center}
\end{figure}

Effects of the $\Lambda NN$ force should instead not be neglected in the Hamiltonian. 
The lowest order in the $\Lambda$-nucleon interaction involves the exchange of two pions (TPE), in
the two- as well as three-body channel (see Fig.~\ref{fig:LN-LNN}). This
is due to isospin conservation which allows only $\Lambda\pi\Sigma$
vertex and so forbids the one pion exchange process.

The $\Lambda N$ potential is written as a sum of a central term, which
also includes a $\Lambda$-N exchange accounting for the one kaon exchange
process, and a spin-dependent part. The TPE process is modeled with
the usual one-pion exchange potential acting twice, as in the Argonne forces.
More details can be found in Refs.~\cite{Usmani:1995,Usmani:2008}.

The three-body $\Lambda$-nucleon interaction takes contributions from
the TPE diagrams in the $s$-wave channel ($v_{\Lambda ij}^{S}$ , Fig.~\ref{fig:LNN_sw})
and in the $p$-wave channel ($v_{\Lambda ij}^{P}$ , Fig.~\ref{fig:LNN_pw}).
Short range effects are accounted for by
a phenomenological central repulsive term $v_{\Lambda ij}^{D}$ :
\begin{eqnarray}
v_{\Lambda ij}^{S}&=&C^{S}Z\left(m_{\pi}r_{i\Lambda\phantom{j}\!\!\!}\right)Z\left(m_{\pi}r_{j\Lambda}\right)
\left({\bm\sigma}_{i}\cdot\hat{\bm r}_{i\Lambda}\;
{\bm\sigma}_{j}\cdot\hat{\bm r}_{j\Lambda}\right){\bm\tau}_{i}\cdot{\bm\tau}_{j} \,,
\nonumber\\[0.3em]
v_{\Lambda ij}^{P}&=&-\left(\frac{C^{P}}{6}\right)\left({\bm\tau}_{i}\cdot{\bm\tau}_{j}\right)
\Bigl\{ X_{i\Lambda}\,,X_{\Lambda j}\Bigr\}\label{eq:V_YNN} \,,
\nonumber\\[0.3em]
v_{\Lambda ij}^{D}&=&W^{D}T_{\pi}^{2}\left(m_{\pi}r_{i\Lambda\phantom{j}\!\!\!}\right)T^{2}_{\pi}\left(m_{\pi}r_{j\Lambda}\right)
\left[1+\frac{1}{6}{\bm\sigma}_\Lambda\cdot \left({\bm\sigma}_{i}+{\bm\sigma}_{j}\right)\right] \,.
\end{eqnarray}
The definition of the functions $Z(x)$ and $X_{i\Lambda}$ 
as well as the set of parameters of the two-body $\Lambda N$ and three-body 
$\Lambda NN$ interactions can be found in Refs.~\cite{Usmani:1995,Usmani:2008}.

We solve the ground-state of the many-body nuclear and hypernuclear
Hamiltonians by means of the AFDMC method, originally introduced by
Schmidt and Fantoni \cite{Schmidt:1999}.
By sampling configurations of the system in
coordinate-spin-isospin space using Monte Carlo algorithms, we
evolve an initial trial wave function $\Psi_T$ in imaginary-time. In the
$\tau\rightarrow\infty$ limit, the evolved state approaches the ground-state
of $H$. Expectation values are computed averaging over the sampled
configurations, and for large $\tau$, the ground
state energy of the system is obtained. For more details see e.g.
Refs.~\cite{Gandolfi:2007,Gandolfi:2009}. 
The extension of the AFDMC method
to $\Lambda$-hypernuclei is technically straightforward.

\section{Results and discussion}
\label{sec:results}
In Tab.~\ref{tab:He5L} the results for $B_\Lambda$ in $^5_\Lambda$He are reported.
Results were obtained using different nuclear
Hamiltonians. The second column corresponds to calculations using the
$\Lambda N$ interaction only. The numerical results with the AV4' and Minnesota
$NN$ potentials are partially consistent. The value obtained with AV6' has a statistically
significant deviation.
All results for $B_\Lambda$ are overestimated, being roughly twice the experimental value.
The third column refers instead to calculations in which the three-body $\Lambda NN$
force, with the set of parameters described in Ref.~\cite{Usmani:1995},
was included. The three-body force significantly reduces the $\Lambda$-separation
energy, yielding a more realistic value.
It is very interesting to notice that the disagreement between  calculations performed using
different nuclear interactions is also reduced below statistical significance.
The same conclusions were obtained by studying
the $^{17}_{~\Lambda}$O hypernucleus, for which the discrepancy between
$B_\Lambda$ computed using different nuclear interactions is few
percent. For this reason we are confident that the $\Lambda$-separation
energy is not too sensitive to the details of the nuclear interaction.

\renewcommand{\arraystretch}{1.2}
\begin{table}[!ht]
\begin{center}
\begin{tabular*}{0.6\linewidth}{@{\hspace{0.5em}\extracolsep{\fill}}lccc@{\extracolsep{\fill}\hspace{0.5em}}}
\midrule
\midrule
Nuclear potential & $B_\Lambda^{\,2B}$	& $B_\Lambda^{\,2B+3B}$	& $B_\Lambda^{\,exp}$ \\
\midrule
Argonne V4'\      & 7.1(1)              &  5.0(1)               & \multirow{3}{*}{3.12(2)} \\
Argonne V6'\      & 6.3(1)              &  5.1(1)               & \\
Minnesota\        & 7.4(1)              &  5.0(1)               & \\
\midrule
\midrule
\end{tabular*}
\caption{$\Lambda$-separation energies (in MeV) for
$^5_\Lambda$He obtained using different nuclear potentials. $B_\Lambda^{\,2B}$
refers to the $\Lambda$-separation obtained with the two-body $\Lambda N$
interaction only. $B_\Lambda^{\,2B+3B}$ is the result for $B_\Lambda$
when the three-body $\Lambda NN$ interaction is included. In the last
column the experimental value of $B_\Lambda$ is from Ref.~\cite{Juria:1973}.}
\label{tab:He5L}
\end{center}
\end{table}

\begin{figure}[!ht]
	\begin{center}
		\includegraphics[width=0.7\linewidth]{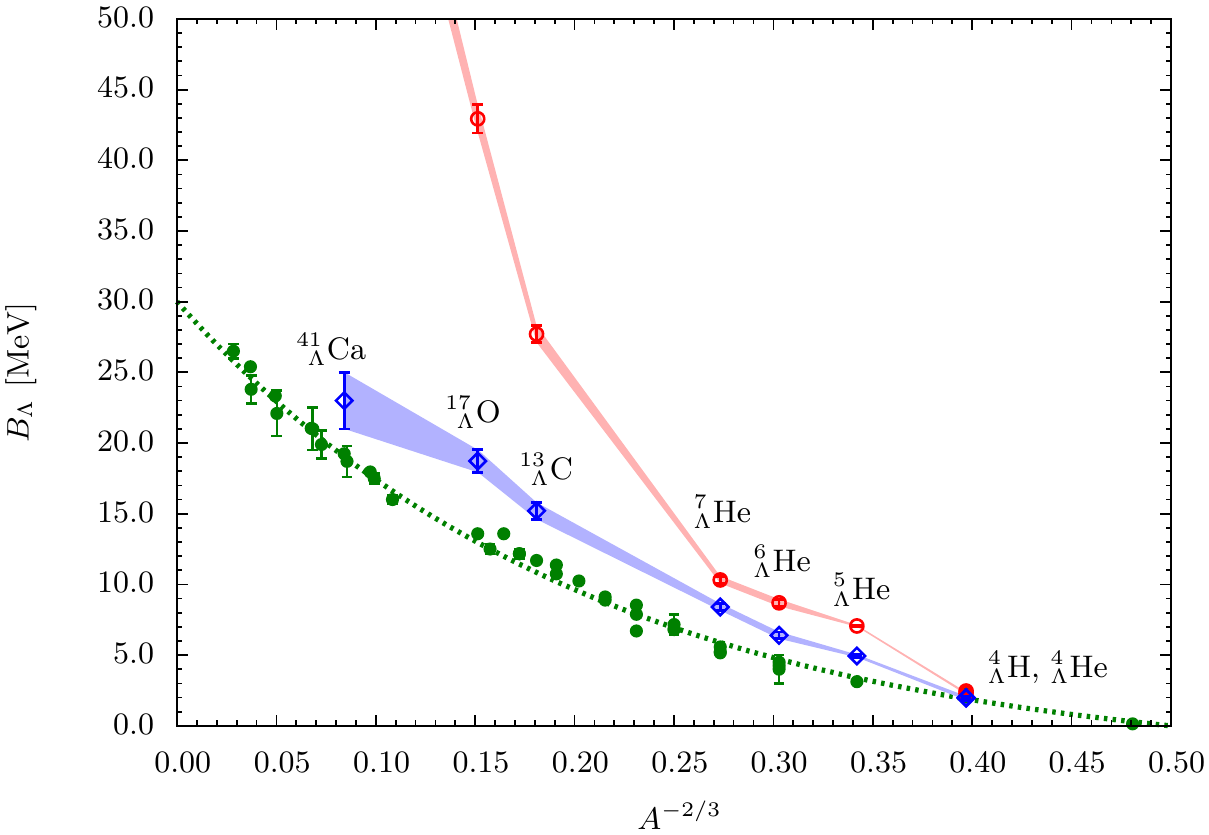}
		\caption{$\Lambda$-separation energy as a function of $A^{-2/3}$. 
		Red circles refer to the AFDMC results for the nuclear AV4' 
		potential plus the two-body $\Lambda N$ interaction. Blue diamonds
		are the results obtained with the same core nuclear
		potential but with both the two- and three- body
		$\Lambda$-nucleon forces. Green solid circles refer to
		the experimental values.}
		\label{fig:BL-A}
	\end{center}
\end{figure}

The $\Lambda$-separation energy as a function of $A^{-2/3}$ is shown in
Fig.~\ref{fig:BL-A}. $B_\Lambda$ is calculated using the two-body $\Lambda
N$ potentials only (upper red curve) and including also the three-body
$\Lambda NN$ interaction (central blue curve). The nuclear potential
is the AV4'. Qualitatively similar results were obtained using the AV6' or Minnesota
$NN$ forces. 
The dashed green curve is the guide line following the experimental data.
It is evident that the two-body $\Lambda$-nucleon interaction alone gives
an unrealistic separation energy. 
The saturation energy, which is estimated to be
around 30~MeV, cannot be reproduced. The inclusion of the three-body
hyperon-nucleon interaction drastically improves the theoretical 
prediction in the separation energies, and
the extrapolated saturation energy becomes reasonably close to experimental data. 
We should point out that a fine tuning of the $\Lambda NN$
parameters might further improve the AFDMC results. In particular, following
Ref.~\cite{Usmani:1995} the $s$-wave channel has not been included in
the present calculations. Preliminary results on $^5_\Lambda$He
and $^{17}_{~\Lambda}$O reveal that this contribution is indeed sub-leading,
and not scaling with the baryon number. The leading term
in the $\Lambda NN$ potential is the repulsive $v_{\Lambda ij}^{D}$,
which is strictly dependent on $A$. Calculations for $^{17}_{~\Lambda}$O
performed with different values of $W^D$, show that it is possible to
move the $\Lambda$-separation energy closer to the expected result of
$13.59$~MeV~\cite{Lalazissis:1988_He5L}, and have a better description of
the $^5_\Lambda$He hypernucleus.

From these results it is clear that the role of the $\Lambda NN$ force
is important in determining the correct $\Lambda$ separation energy,
and should not be neglected. The fact that the leading contribution to the
three-body interaction is strictly repulsive in the range of hypernuclei studied,
and the fact that it scales with the number of baryons, suggests that
the $\Lambda$-nucleon potential discussed in this paper, when applied
the study of the homogeneous medium, should lead to a
stiffer EOS for the $\Lambda$-neutron matter.
This fact might eventually reconcile the onset of hyperons in the inner core of a 
NS with the observed masses of order $2~M_\odot$. 
A study along this direction is in progress.

\section*{Acknowledgments}
\pdfbookmark[0]{Acknowledgments}{Acknowledgments}
\label{sec:acknowledgments}
Computer time was made available by the AuroraScience project (funded
by PAT and INFN) in Trento, by Los Alamos Open Supercomputing, and by
the National Energy Research Scientific Computing Center (NERSC).
The work of S.~G. is supported by the Nuclear Physics program
at the DOE Office of Science, UNEDF and NUCLEI SciDAC programs,
and by the LANL LDRD program.

\bibliographystyle{model1a-num-names}

\end{document}